\documentclass[11pt]{article}    
\usepackage{amsmath,amssymb,amsfonts} 
\usepackage{graphics}                 
\usepackage{color}                    
\usepackage{amsmath}
\usepackage{cancel}
\usepackage{geometry}                
\newcommand{\be}{\begin{eqnarray}}
\newcommand{\en}{\end{eqnarray}}

\newcommand{\pa}[1]{\partial_{#1}}

\DeclareMathAlphabet{\mathpzc}{OT1}{pzc}{m}{it}
\title{Automorphism Induced  Nonlocal Conservation Laws }

\author{Clifford Chafin\\\ \small{Department of Physics, North Carolina State University, Raleigh, NC 27695} \thanks{cechafin@ncsu.edu}}

\begin{document}
\maketitle
\begin{abstract}
The conservation laws of electromagnetism, and implicitly all theories built from quadratic Lagrangians, are extended to a continuum of nonlocal versions.  These are associated with symmetries of a class of equal time field correlation functions and give results for both connected and disconnected branches of the general linear group of the space.  
It is generally assumed that manifestly covariant Lagrangians are the necessary starting point for physical theories.  Here we show that the EOM derived from any of these can also follow from a broad class of nonlocal ones and each generally gives a different nonlocal Noether current.  When the equations are put into a linear form and evaluated on a flat spacetime, a simple ansatz exists to give a class of conservation laws corresponding to all affine transformations of the underlying space.  A general procedure is given to generate a class of nonlocal conservation laws for solutions to a very large class of  nonlinear PDEs.  
\end{abstract}

Conservation laws play a central role in physics.  They allow us to solve a broader class of problems than spatial symmetry can alone where the detailed dynamics can be quite complicated.  Additionally, they give us a set of notions on which to build intuition, place bounds on physical behavior and relate dynamics across systems that behave according to different macroscopic laws.  In special cases, they are the key to complete exact solutions.  This article is built on a set of observations that suggest conservation laws can be much more general and often less ``physically'' meaningful than are usually assumed.  

In graduate school, while studying electrodynamics and playing with vector calculus identities, I found a conservation law that involved a density that is the product of quantities at two different locations.  
\begin{equation}
\rho(x^{i},t)={B}(x^{i},t) \cdot {E}(-x^{i},t) +  {B}(-x^{i},t) \cdot {E}(x^{i},t)\} 
\end{equation}
(in contrast to the local energy density $\mathcal{E}=E^{2}+B^{2}$).
This ``2-point'' conservation law does not seem to follow from any symmetry of the Lagrangian.  Its physical importance seemed doubtful to me since locality is fundamental to causality in nature yet there turned out to be a whole family of such conservation laws; all inequivalent to each other.  We will see that there are ways to rewrite the Lagrangian in terms of nonlocal expressions that give the same local equations of motion.  These new ``2-point Lagrangians'' can possess a set of continuous symmetries to which Noether's Theorem applies.  This procedure is robust enough to extend to many other Lagrangians and can be extended in many ways to give a whole continuum of conservations laws.  

Given a Lagrangian with a set of $n$ continuous symmetries, Noether's theorem guarantees $n$ corresponding conservation laws.  The stress energy tensor gives ten of these in flat spacetime.  The local conservation laws for these are of the form $\nabla \cdot j = \partial_t \rho$  where $\rho=T_{i0}K^{i}$ where $K^{i}$ is a Killing vector of the space.  To each of these local conservation laws there is a global version:  
\begin{equation}
\dot{Q}=\frac{d}{dt}\int \rho ~d^{3}x = 0
\end{equation}
or
\begin{equation}
\dot{Q}=\frac{d}{dt}\int \rho ~d^{3}x = \mbox{Rate of Source Contributions}  
\end{equation}  
when there are sources.  These global conservation laws depend on the spatial symmetries and separability of the time direction.  Some of these, like the system net momentum, become vectorial quantities under the global boost transformations of the space.  

These conservation laws can be found by Noether's result that each continuous symmetry of the Lagrangian generates a conservation law.  This result and the invariance properties of special relativity, often with the later insights of general relativity, dominate physicists' thinking on conservation laws and their meaning.  There have been interesting developments in the study of the KdV equation and role of an infinite set of invariants on closed form integrability of a system.  Results from studies of the jet space structure an internal symmetries of differential systems has led to a number of nonlocal conservation laws for particular wave equations and other partial differential systems \cite{Anco, Vinogradov}.  These can be thought of as ``nonlocal'' in the sense that any infinite order expression can be interpreted as a power series of an analytic function and so evaluated at a point at some finite distance away.  

In one sense, nonlocal conservation laws typically exist.  If one allows the form of the conservation law to be a function of the particular solution and a PDE that is linear and first order in time then one can construct it trivially from data on a finite subset of points.  Consider a solution to the operator equation $\hat{\mathcal{O}}f(x,t)=\partial_{t}f+\hat{\mathcal{O}}_{x}f=0$ on a compact interval $I$.  Assume that the set of functions $\partial_{x}^{n}f|_{t=0}$ separate points in the sense of the Stone-Weierstrass theorem \cite{Conway}.  Select a finite subset of points in the interval $x_{i}\in I$ so that the values $\hat{\mathcal{O}}_{x}^{n}f|_{x_{i}}$ all differ.  By linear combinations we can construct functions $g_{n}=\sum_{m}\alpha_{ni}x_{i}$ so that $\partial_{t}^{n}g_{m}=0$ for all $n, m$.  This ensures that each $g_{n}(f)$ is an invariant function.  Even though this is a discrete example one can take limits to a dense set of points on the space.  The question of when the coefficient functions can be chosen as continuous or $\mathcal{C}^{\infty}$ is interesting but not to our purpose here.  The operator $\hat{\mathcal{O}}_{x}$ can be higher order and nonlinear.  This shows that nonlocal invariants are easy to construct even in the nonlinear case for a very broad class of operators and solutions.  

The case of integrability depends on the existence of an infinite set of invariants which have a form independent of the particular solution given.  For ODEs, conservation laws induce foliations of the solution space and, given enough of them, produce solutions that are not plagued by the chaotic folding problems that characterize the Lorentz attractor so that solutions with different initial data maintain some ``reasonable'' separation over time \cite{Hirsch, Abraham}.  For the sake of early disclosure, these are not the kinds of conservation laws we are generating here.  It will be hard to assign ``physical'' meaning to them but, given that the details of dynamical evolution are usually hard, bounding constraints can be very powerful so we adopt the attitude that ``any conservation law is a good conservation law.''

Of course, one can find conservation laws when no Lagrangian exists.  Given a set of linear homogeneous PDEs in a set of vector fields 
\begin{align}
\partial_{t}F_{i}^{(r)}=M_{ijk}^{(rs)}\partial_{j}F_{k}^{(s)}+J^{(r)}_{i}
\end{align}
where the indices in parenthesis are labels of the field type and the others are rectangular coordinate indices and $J$ indicates driving currents.\footnote{The summation convention is assumed on repeated indices and field type labels here.}  We can seek a quadratic scalar conserved quantity $\mathpzc{P}(F)=W_{ij}^{(rs)}F_{i}^{(r)}F_{j}^{(s)}$ with some position independent constants $W_{ij}^{(rs)}$.  Specifically we desire a current $\mathcal{J}_{i}(F_{j}^{(r)})=K_{ijk}^{(rs)}F_{i}^{(r)}F_{j}^{(r)}$ such that $\partial_{t}\mathpzc{P}(F)+\nabla_{(x)}\cdot\mathcal{J}(F)=0$ or $\partial_{t}\mathpzc{P}(F)+\nabla_{(x)}\cdot\mathcal{J}(\partial_{i}F)=S(J,F)$, some source term in terms of the driving currents.  The usual example of this is the electromagnetic energy density $\mathcal{E}_{em}$ and Poynting vector $\mathcal{P}_{i}$ \cite{Jackson}.\footnote{We now see that we have a bit of an inconvenience in terminology.  We need a notion of charge current that acts as sources for the fields and a notion of ``momentum'' current that acts as a source of field ``energy.''  The scripted notation will be reserved for the case of current of conserved quantities.}  
Since the equations are linear and homogeneous, the task is then to eliminate cross terms by use of the EOM and a nontrivial choice of $W_{ij}^{(rs)}$ and $K_{ijk}^{(rs)}$.  It will be seen that nothing forces us to use a local prescription for this task.  We can seek a ``two-point'' density of the form $\mathpzc{P}(F)(x,x')=W_{ij}^{(rs)}F_{i}^{(r)}(x)F_{j}^{(s)}(x')$ and use some automorphism $\hat{A}(x)
=x'$ so that our new density can be expresses as a local function $\mathpzc{P}(F)(x,x')\rightarrow \mathpzc{P}(F)(x,\hat{A}x)$.  When $\hat{A}$ is an affine transformation, $A_{ij}x_{i}+b_{j}$, the derivatives give a simple form that will often be amenable to our task of forming new conserved densities and currents.  Since we have inserted a spatial automorphism into the transformation, it is not reasonable to call this a local function even though it is manifestly a function of only $x$ and $t$.  For cases that do follow from the action principle, let us now look at this situation from an Lagrangian point of view.

In the cases where we seek to derive conservation laws, we are interested in finding a set of nonlocal Lagrangians that generate the same EOM.  It will turn out that there are many of these.  If the EOM is given by $\hat{\mathcal{O}}f=0$, then graphically we write
$\mathcal{L}_{\hat{\mathcal{O}}}\rightarrow [\hat{\mathcal{O}}f=0] \leftrightarrow \{\mathcal{L}_{\hat{\mathcal{O}}_{i}}\}$ to denote the class of $i$-indexed Lagrangians that specify the same local EOM.  This task is significantly simpler if we let the sources $J$ be externally driven and not have to respond to the fields and conserve energy and momentum.  However, in the next section we will include the current completely generally for the case of Dirac field.  The key to this program lies in introducing auxiliary fields that obey a propagated set of constraints at a distant point as $A(x)=\tilde{A}(-x)$ and $\tilde\psi(-x)=\psi(x)$.  The tilde denotes an independent field and one then has the option of considering it as a field at a distant argument or redefine it as a new local field $F(x)=\tilde{A}(-x)$ in the variable $x$.  When the point transformations are more complicated we will see that the first point of view is more illuminating.

\section{Two-Point Lagrangians}
The extremal action condition $\delta\int \mathcal{L}=0$ gives the equations of motion where, by the variation, we generally mean an arbitrary change in the field functions themselves.  Quadratic homogeneous terms in the field gradients give second order equations.  In the case of electromagnetism the action is generally written as $\mathcal{L}=-(4\mu_{0})^{-1}F_{\mu\nu}F^{\mu\nu}$ where $F_{\mu\nu}=\partial_{[\mu}A_{\nu]}=\partial_{\mu}A_{\nu}-\partial_{\nu}A_{\mu}$ expresses the action in terms of the vector potential $A_{\mu}$.  Coupling this gauge dependent quantity to a (gauge invariant) current $J$ in $\mathcal{L}_{int}=A_{\mu}J^{\mu}$ introduces some  inconsistency unless $\mathcal{L}_{J}$ is chosen to have a gauge dependence that cancels it.  In the case of the Dirac Lagrangian, $\mathcal{L}_{J}(\psi)=i\hbar c\psi^{\dagger}\gamma^{0}\gamma^{\mu}\partial_{\mu}\psi-mc^{2}\psi^{\dagger}\gamma^{0}\psi$, this term appears as the gradient of a phase term $\partial_{\mu}\chi$.\footnote{A detailed discussion on implied gauge choices for the classical particle Lagrangian is given in the appendix of \cite{Chafin-EM}.}  

Assuming $F_{\mu\nu}$ is an implicit function of $A_{\mu}$, the usual interacting Lagrangian for electrodynamics is given by 
\begin{align}\label{LF}
\mathcal{L}=-\frac{1}{4\mu_{0}}F_{\mu\nu}F^{\mu\nu}+A_{\mu}J^{\mu}+\mathcal{L}_{J}
\end{align}  
Variation with respect to $A_{\mu}$ give the EOM of the fields as $\partial_{\mu}F^{\mu\nu}=-\mu_{0}J^{\mu}$ where the homogeneous Maxwell equations follows from the antisymmetry of $F^{\mu\nu}$.  This induces a second order equation on $A^{\mu}$ but, since the equations can be written manifestly with out it, any such $A^{\mu}$ fields that give the same $F^{\mu\nu}$ are equivalent.  
It is often said that the Maxwell equations do not follow from a Lagrangian but this is not entirely true.  Assuming an antisymmetric field $C_{\mu\nu}$ corresponding to the electromagnetic fields $F_{\mu\nu}$ we can derive a set of equations of motion for the pair $C_{\mu\nu}, A_{\rho}$ from
\begin{align}\label{LC}
\mathcal{L}=-\frac{1}{\mu_{0}}(2^{-1}C_{\mu\nu}C^{\mu\nu}+C^{\mu\nu}\partial_{\mu}A_{\nu})+A_{\mu}J^{\mu}+\mathcal{L}_{J}
\end{align}  
These give first order equations $\partial_{\mu}C^{\mu\nu}=-\mu_{0}J^{\mu}$ and the constraint $C_{\mu\nu}=\partial_{[\mu}A_{\nu]}$ under 
variation by $C_{\mu\nu}$ and $A_{\nu}$ respectively.  

Let us now consider modifications of these Lagrangians under the simple coordinate transformation $x^{i}\rightarrow\mathcal{F}(x^{i})=-x^{i}$ of some of the functions in the quadratic pairs.  Our goal is to generate the same equations of motion as before.  This is prohibitive with either of the Lagrangians in the form eqn\ \ref{LF} or eqn\ \ref{LC}.  The resolution is to utilize a set of auxiliary fields $\tilde{C}_{\mu\nu}, \tilde{A}_{\mu}, \tilde{\psi}$ that will be varied independently but later chosen to be related to their associated tilde-free components in applications for physical initial data.  To this end, consider the bilinear (rather than quadratic) Lagrangian
\begin{align}\label{LCDual}
\mathcal{L}_{\text{bilin}}(x,t;\mathcal{F})=&-\frac{1}{\mu_{0}}\bigg(C_{\mu\nu}(x,t)\tilde{C}^{\mu\nu}(-x,t)\\\nonumber
&+\tilde C^{\mu\nu}(-x,t)\partial_{\mu}A_{\nu}(x,t))+ C^{\mu\nu}(x,t)\partial_{\mu}\tilde A_{\nu}(-x,t)\bigg)\\\nonumber
&+A_{\mu}(x,t) J^{\mu}(\tilde\psi,-x,t)+\tilde A_{\mu}(-x,t)J^{\mu}(\psi,x,t)\\\nonumber
&+\mathcal{L}_{J}(\psi(x,t))+\mathcal{L}_{ J}(\tilde \psi(-x,t))
\end{align} 
Varying these six fields $C_{\mu\nu},A_{\mu},\psi,\tilde{C}_{\mu\nu}, \tilde{A}_{\mu},\tilde\psi$, gives the EOM $\partial_{\mu}C^{\mu\nu}=-\mu_{0}J^{\mu}$, the constraint $C_{\mu\nu}=\partial_{[\mu}A_{\nu]}$ and the Dirac equation $i\hbar c\gamma^{\mu}(\partial_{\mu}+ieA_{\mu})\psi-mc^{2}\psi=0$ and the same equations for the tilde fields evaluated at $(-x,t)$ provided we choose initial data $\tilde A_{\mu}(x,t)=A_{\mu}(x,t)$, $\tilde{\psi}(x)=\psi(-x)$.  We see that this condition is propagated and the nonlocal Lagrangian eqn~\ref{LCDual} generates the desired EOM.\footnote{For a more general take on such bilinear Lagrangians that incorporates GR and a proof that they always give causality of gauge invariant quantities see \cite{Chafin-Dual}.}  Noether's Theorem generates a conserved stress-energy tensor despite the nonlocal information included.  A symmetric tensor can be found by the usual procedure of converting to covariant derivatives and varying $\mathcal{L}\sqrt{g}$ with respect to $g_{\mu\nu}$.  

We can generalize this Lagrangian in a couple of ways.  Firstly, we can leave the spatial transformation the same and introduce a (nondegenerate) constant mixing of the fields that preserves the form of the equations of motion.  Specifically, define
\begin{align}
\mathcal{L}_{\text{mix}}(x,t;\mathcal{F})=&-\frac{1}{\mu_{0}}g^{\mu\sigma}M^{\nu\rho}\bigg(C_{\mu\nu}(x,t)\tilde{C}_{\sigma\rho}(-x,t)\\\nonumber
&+\tilde C_{\mu\nu}(-x,t)\partial_{\sigma}A_{\rho}(x,t))+ C_{\mu\nu}(x,t)\partial_{\sigma}\tilde A_{\rho}(-x,t)\bigg)\\\nonumber
&+g^{\sigma\rho}\bigg(A_{\rho}(x,t) J_{\sigma}(\tilde\psi,-x,t)+\tilde A_{\rho}(-x,t)J_{\sigma}(\psi,x,t)\bigg)\\\nonumber
&+\mathcal{L}_{J}(\psi(x,t))+\mathcal{L}_{ J}(\tilde \psi(-x,t))
\end{align} 
where 
\begin{align}
J^{\mu}&=i\hbar c\psi^{'\dagger}\gamma^{0}M^{\mu}_{\nu}\gamma^{\nu}\psi'=i\hbar c\psi^{\dagger}\gamma^{0}\gamma^{\mu}\psi\\
\psi'&=S(M^{-1})\psi
\end{align}
This $\psi'$ is the result of an active transformation of the original $\psi$ to give the correct identities.  In this sense we have changed the system but derived a corresponding conservation law for another current.  

In this formula $S(M)$ is the (not necessarily unitary) matrix transformation corresponding to the matrix transformation $M_{\mu\nu}$ of the vector indices.  In the unitary case, $M_{\mu}^{\nu}\leftrightarrow\Lambda_{\mu}^{\nu'}$, it can be associated with an active boost and rotation of the wavefunction $\psi$ \cite{Bjorken}.  Since this is a purely spatial transformation, it commutes with the matrix $\gamma^{0}$ and the evolution commutes with the mapping.

We can also generalize the inversion map to any affine transformation, $\mathcal{G}:x_{i}\rightarrow\alpha_{ij}x_{i}+\beta_{j}$, ($x'=\alpha x+\beta$).  This requires some additional modification of the Lagrangian since the transformation is not generally an isometry.
\begin{align}
\mathcal{L}_{\text{affine}}(x,t;\mathcal{G})=&-\frac{1}{\mu_{0}}\bigg(C_{\mu\nu}(x,t)\tilde{C}^{\mu\nu}(\alpha x+\beta,t)\\\nonumber
&+\tilde C^{\mu\nu}(\alpha x+\beta,t) \partial_{\mu}A_{\nu}(x,t))+ C^{\mu\nu}(x,t)\alpha^{-1}_{\mu\sigma}\partial^{\sigma}\tilde A_{\nu}(\alpha x+\beta,t)\bigg)\\\nonumber
&+ A^{\mu}(x,t) \alpha_{\mu\nu} J^{\nu}(\tilde\psi,\alpha x+\beta,t)+\tilde A^{\mu}(\alpha x+\beta,t) \alpha^{-1}_{\mu\nu} J^{\nu}(\psi,x,t)\\\nonumber
&+\mathcal{L}_{J}(\psi(x,t))+\mathcal{L}_{ J}(\tilde \psi(\alpha x+\beta,t))
\end{align} 
The matrix $\alpha$ is now extended to a four space version by embedding in Lorentz space as $\alpha_{ij}\rightarrow \alpha_{\mu\nu}=$ diag$(0,\alpha_{ij})$.\footnote{One could argue that the action should now include a measure change of $|\alpha|^{-1/2}$ however, since it is a constant it makes no change in the evolution or conservation laws.}  Variation of the action by $\tilde A$ gives $\nabla_{(x)}\cdot C(x)=J(x)$ and $A$ gives $(\alpha^{-1}\nabla_{(x)})\cdot C(\alpha x+\beta)=\nabla_{(\alpha x+\beta)}\cdot C(\alpha x+\beta)=J(\alpha x+\beta)$.  Variation by $\tilde C$ gives $C_{\mu\nu}(x)=\nabla_{(x) [\mu}A_{\nu]}(x)$ and by $C$ yields $\tilde C_{\mu\nu}(\alpha x+\beta)=\nabla_{(\alpha x+\beta) [\mu}\tilde A_{\nu]}(\alpha x+\beta)$.  If we assume initial data such that $\tilde A(\alpha x+\beta)=\alpha A(x)$ and $\tilde J(\alpha x+\beta)=\alpha J(x)$ (so that $\tilde \psi(\alpha x+\beta)= S \psi(x)$ with $S^{\dagger}\gamma^{\mu}S=\alpha^{\mu}_{\nu}\gamma^{\nu}$) then these conditions are propagated and the usual local EOM result.

\section{Specific Examples from Maxwell's Equations}
\subsection{Inversion}

As an example of the method of deriving conservation law directly from eom we consider the inversion map $\mathcal{F}:x\rightarrow -x$.  The usual local energy conservation law for Maxwell's equations can be written:
\begin{eqnarray}
\partial_{t} ({E}\cdot {E} + {B}\cdot {B}) +2 \nabla  \cdot   ({E} \times {B}) =  - 2{J} \cdot {E}\\
\Rightarrow   \partial_{t}\int ({E}\cdot {E} + {B}\cdot {B}) d^{3}x = -2 \int {J} \cdot {E} d^{3}x
\end{eqnarray} 
where we have assumed the fields vanish sufficiently rapidly for the integrals to be finite and the divergence term to be reduced to a vanishing surface term.  This, like all the usual conservation laws of electromagnetism, can be found by using one of the ten global symmetries of the Lagrangian $-\frac{1}{4}F^{\mu \nu} F_{\mu \nu} +J^{\alpha}A_{\alpha} + \mathcal{L}_{J}$.

The nonlocal result associated with the inversion map we will prove is
\begin{align}
\nabla_{(x)} \cdot 
\{
{B}(x) \times {B}(-x)) - ({E}(x) \times {E}(-x) \} =& \partial_{t} \{ {B}(x) \cdot {E}(-x)+  {B}(-x) \cdot {E}(x)\}\\ &
+ \{{B}(x)\cdot {J}(-x) + {B}(-x)\cdot {J}(x)\}
\label{exoticlaw}
\end{align}
or expressed in global form
\begin{equation}
\partial_{t} \int \{{B}(x) \cdot {E}(-x) +  {B}(-x) \cdot {E}(x)\} dx^{3} + 
\int \{{B}(x)\cdot {J}(-x) + {B}(-x)\cdot {J}(x)\} dx^{3} = 0
\end{equation}
where $x$ here refers only to the spatial coordinates and $\nabla_{(x)}$ mean the divergence w.r.t.\ the variable $x$.  (The time label has been suppressed).  This looks much like the usual energy conservation law with the external driving term given by ${E}\cdot {J}$ exchanged with ${B}\cdot {J}$.  This equation can be verified by taking derivatives $\partial_{x}$ and interpreting expressions like $\partial_{x}E(-x)$ as $\partial_{x}E(\mathcal{F}(x))$. 

We can prove this relation directly 
by considering the difference of the two expressions \mbox{$\nabla_{(x)} \cdot (E(x) \times E(-x))$} and 
\mbox{$\nabla_{(x)} \cdot (B(x) \times B(-x))$} and seeking a ``two-point'' identity similar to the usual identities of vector calculus.
Using Maxwell's equations on the first gives:
\begin{align*}
\nabla_{(x)} \cdot (E(x) \times E(-x)) =& E(-x) \cdot \nabla_{(x)} \times E(x) -  
E(x) \cdot \nabla_{(x)} \times E(-x) \nonumber \\
=& E(-x) \cdot \nabla_{(x)} \times E(x) +  
E(x) \cdot \nabla_{(-x)} \times E(-x)  \nonumber \\
=& E(-x) \cdot (-\partial_{t} B(x)) +  
E(x) \cdot (-\partial_{t} B(-x)) \nonumber
\end{align*}
and on the second:
\begin{align*}
\nabla_{(x)} \cdot (B(x) \times B(-x)) =& B(-x) \cdot \nabla_{(x)} \times B(x) -  
B(x) \cdot \nabla_{(x)} \times B(-x)
\nonumber \\
=& B(-x) \cdot \nabla_{(x)} \times B(x) +  
B(x) \cdot \nabla_{(-x)} \times B(-x)
\nonumber \\
=& B(-x) \cdot (J(x) + \partial_{t} E(x)) +  
B(x) \cdot (J(-x) + \partial_{t} E(-x))
\nonumber \\
=& \{ B(-x) \cdot J(x) + B(x) \cdot J(-x) \} +  
\{ B(-x) \cdot \partial_{t} E(x)) +  B(x) \cdot \partial_{t} E(-x) \} \nonumber
\end{align*}
Combining these two equations \\ \\
$\nabla_{(x)} \cdot 
\{
B(x) \times B(-x) - E(x) \times E(-x) \} = \\ \\
~~~~~ ~~~~~~~~~\partial_{t} \{ B(x) \cdot E(-x) +  B(-x) \cdot E(x)\} + 
\{B(x)\cdot J(-x) + B(-x)\cdot J(x)\}$\\

Assuming the fields vanish at infinity we can integrate and eliminate the divergence to obtain:
\begin{equation}
\partial_{t} \int \{B(x) \cdot E(-x) +  B(-x) \cdot E(x)\} dx^{3} + 
\int \{B(x)\cdot J(-x) + B(-x)\cdot J(x)\} dx^{3} = 0.
\end{equation}
We can refer to the quantity $B(x)\cdot J(-x) + B(-x)\cdot J(x)$ as the ``3-space pseudoscalar energy associated with the 2-point spatial inversion map.''  For contrast we next consider the case of rotations.

\subsection{Rotations}
Consider a transformation $x'^{j}=R^{i}_{~j}x^{j}$ ($x'=Rx$).  We investigate the divergence of the expression $\mathcal{J}:=E(x)\times RB(Rx)+RE(Rx)\times B(x)$.  We will need an identity for the cross product under rotations.  Clearly $Rx\times Ry=R(x\times y)$.  This induces the identity $\epsilon^{ijk}R^{l}_{j}R^{m}_{k}=\epsilon^{plm}R^{i}_{p}$.  Therefore $R(\nabla_{(x)}\times v)=\nabla_{(R^{-1}x)}\times Rv$ or $\nabla_{(x)}\times v=R(\nabla_{(Rx)}\times R^{-1}v)$.
\begin{align}
\nabla\cdot \mathcal{J} &=\nabla\cdot\big\lbrace  E(x)\times RB(Rx)+RE(Rx)\times B(x) \big\rbrace\\\nonumber
&=\big\lbrace  RB(Rx)\cdot\nabla\times E(x)-  E(x)\cdot\nabla\times RB(Rx)\\\nonumber
&~~~~~~~~~~~~~+B(x)\cdot\nabla\times RE(Rx)-  RE(Rx)\cdot\nabla\times B(x)   \big\rbrace\\ \nonumber
&=\big\lbrace  RB(Rx)\cdot\nabla\times E(x)-  E(x)\cdot R(\nabla_{(Rx)}\times B(Rx)) \\\nonumber
&~~~~~~~~~~~~~+B(x)\cdot R(\nabla_{(Rx)}\times E(Rx))-  RE(Rx)\cdot\nabla\times B(x)   \big\rbrace\\ \nonumber
&=\big\lbrace  RB(Rx)\cdot(-\pa{t}B(x))
-  E(x)\cdot R(J(Rx)+\pa{t}E(Rx)) \\\nonumber
&~~~~~~~~~~~~~+B(x)\cdot R(-\pa{t}B(Rx))
-  RE(Rx)\cdot(J(x)+\pa{t}E(x))   \big\rbrace\\ \nonumber
&=\pa{t}\big\lbrace -B(x)\cdot RB(Rx) -E(x)\cdot RE(Rx) \big\rbrace-\big\lbrace E(x)\cdot RJ(Rx) +RE(Rx)\cdot J(x)  \big\rbrace
\end{align}
which gives us a 2-point energy $\mathpzc{P}=\int B(x)\cdot RB(Rx) +E(x)\cdot RE(Rx) d^{3}x$ which is preserved when the ``2-point rotation work'' done by the current vanishes.

\section{Symmetries for 2-point Time and Space Displacements}
\label{displacements}

The easiest 2-point Lagrangians we can build use discrete translation of some of the time and space coordinates in the Lagrangian.  Fix a pair of displacements $\Delta x$ and $\Delta t$ and consider 
\begin{align*}
\mathcal{L}:=-\frac{1}{4}\{F^{\mu \nu}(x+\Delta x,t+\Delta t)F_{\mu \nu}(x,t)+&J^{\alpha}(x+\Delta x,t+\Delta t)A_{\alpha}(x,t)\\
+F^{\mu \nu}(x,t)F_{\mu \nu}(x+\Delta x,t+\Delta t)+&J^{\alpha}(x,t)A_{\alpha}(x+\Delta x,t+\Delta t)\}.
\end{align*}
where the symmetrization is over the shifted and unshifted coordinates and the currents are externally constrained.

To vary the action we need to consider variations of terms like $\pa{}A\pa{}A$.  Varying $A_{\beta}(x,t)\rightarrow A_{\beta}(x,t)+\delta^{\gamma}_{\beta}\delta(x-y,t-t')$ in action over arbitrary volumes (that may or may not include both points $x$ and $-x$).  Typical variation terms are:
\begin{align*}
\hspace*{-1.5cm}
&{\delta S_{\tiny (\pa{}A\pa{}A)}}={\delta}\int g^{\alpha \mu}g^{\beta \nu}\pa{\alpha}A_{\beta}(x+\Delta x,t+\Delta t)\pa{\mu}A_{\nu}(x,t)dx^{3}dt\\
=&\int \bigg\lbrace g^{\alpha \mu}g^{\beta \nu}\pa{\alpha}[A_{\beta}(x+\Delta x,t+\Delta t)+\delta^{\gamma}_{\beta}\delta(x+\Delta x-y,t+\Delta t-t')]\\
&~~~~~~\times\pa{\mu}[A_{\nu}(x,t)+\delta^{\gamma}_{\nu}\delta(x-y,t-t')]\\
&~~~~~~-g^{\alpha \mu}g^{\beta \nu}\pa{\alpha}A_{\beta}(x+\Delta x,t+\Delta t)\pa{\mu}A_{\nu}(x,t)\bigg\rbrace  dx^{3}dt\\
=&-\int \bigg\lbrace  \delta^{\gamma}_{\nu}\delta(x-y,t-t')g^{\alpha \mu}g^{\beta \nu}\pa{\mu}\pa{\alpha}A_{\beta}(x+\Delta x,t+\Delta t)\\
&~~~~~~~+\delta^{\gamma}_{\beta}\delta(x+\Delta x-y,t+\Delta t-t')g^{\alpha \mu}g^{\beta \nu}\pa{\alpha}\pa{\mu}A_{\nu}(x,t) \bigg \rbrace  dx^{3}dt\\
=&- \bigg\lbrace  g^{\alpha \mu}g^{\beta \gamma}\pa{\mu}\pa{\alpha}A_{\beta}(y+\Delta x,t'+\Delta t)
+g^{\alpha \mu}g^{\gamma \nu}\pa{\alpha}\pa{\mu}A_{\nu}(y-\Delta x,t'-\Delta t) \bigg \rbrace \\
=&- \bigg\lbrace  \Box A^{\gamma}(y+\Delta x,t'+\Delta t)
+\Box A^{\gamma}(y-\Delta x,t'-\Delta t) \bigg \rbrace. \\
\end{align*}
(Lorentz gauge assumed here for simplicity)
where we have written this out in detail because this sort of variation is novel\footnote{The ``$\delta\delta$'' contributions are neglected.  We should really consider delta functions be be replaced by finite sized small amplitude distributions with height much smaller than 1 to make this valid.  The common procedure to define functional differentiation this way, as in QFT by Ryder \cite{Ryder}, seems to overlook this problem.}.
These variations over the sum of such terms will equal $J^{\alpha}$.  The variations give results at different locations so must vanish independently.  This gives the usual (inhomogeneous) Maxwell equations.  Notice that the derivatives $\pa{i}$ were all with respect to $(x,t)$ and that these are the same as being with respect to any of the shifted coordinates.  This is why the variables the partial derivatives acted with respect to could be suppressed throughout the calculation.

Not surprisingly we get the usual equations of motion and conservation laws like\footnote{Analogously to above, the term ${E}(x+\Delta x,t+\Delta t)\cdot{E}(x,t)+{B}(x+\Delta x,t+\Delta t)\cdot{B}(x,t)$ could be labeled the 3-space scalar energy associated with the 2-point translation map.}:
\begin{align}
\partial_{t}\int \{{E}(x+\Delta x,t+&\Delta t)\cdot{E}(x,t)+{B}(x+\Delta x,t+\Delta t)\cdot{B}(x,t)\}d^{3}x\\
& =-\int \{J(x+\Delta x,t+\Delta t)\cdot E(x,t)+ J(x,t)\cdot E(x+\Delta x,t+\Delta t)\} d^{3}x
\end{align}
where the usual conservation laws are obtained by letting $\Delta t, ~\Delta x \rightarrow 0$.

\subsection{Example: Plane EM Waves}

In usual vector calculus notation, the plane electromagnetic wave propagating in the $\hat{z}$ direction is described by:
\begin{align*}
\vec{E}(\vec{x},t)={E}_{0}\sin{(k\hat{z}\cdot \vec{x}-\omega t)}\hat{x}\\
\vec{B}(\vec{x},t)={E}_{0}\sin{(k\hat{z}\cdot \vec{x}-\omega t)}\hat{y}
\end{align*}

Let us consider the above 2-point energy with $\Delta \vec{x} =d\hat{z}$ and $\Delta t =0$.  
\begin{align*}
\mathcal{E}:=&\int \{{E}(x^{i}+d^{i},t)\cdot{E}(x,t)+{B}(x^{i}+d^{i},t)\cdot{B}(x,t)\}d^{3}x\\
=&\int \{2E_{0}^{~2}\sin{(k(z+d)-\omega t)}\sin(kz-\omega t)\}dx dy dz\\
=&(\mbox{Vol})E_{0}^{~2} \cos(kd)
\end{align*}
We see that when $d$ is an integer multiple of the wavelength $\lambda$ we get the usual electromagnetic energy and when it is half-integer the 2-point energy vanishes.

If we choose the size of the box, $L$ and the shift size, $d$ to be multiple of $\lambda$ then we get $\mathcal{E}\equiv(\mbox{Vol})E_{0}^{~2} \cos(kd)$ which is manifestly time independent.  This can be combined with an opposite moving wave that gives a true standing wave with persistent nodes at the walls.  This eliminates surface terms for the cell (and for the cell shifted by $d\hat{z}$) so the 2-point energy is {exactly} conserved.  

\section{Conclusions}
Nonlocal conservation laws are very plentiful in the case of EOM derived from quadratic Lagrangians.  The would seem to be so for many linear sets of equations.  The importance of such laws can be debated.  Since these quantities can be zero in the cases where the usual energy, momentum\ldots vanish they clearly contain information not present in the usual local conservation laws.  Linear equations tend to have bases for solutions so there is no urgency in providing more solutions to them.  Bounds on solutions play important roles in estimating solutions with difficult boundary conditions and in numerical solutions.  We have also seen that there can be plentiful nonlocal conservation laws for nonlinear PDEs provided one does not require the form of these solutions be independent of the particular solution itself.  

One of the most compelling aspects of these laws is their simplicity and they may possibly influence how we perceive the usual conservation laws.  In the case of inhomogeneous (typically nonquadratic) Lagrangians there is no clear path to generating nonlocal alterations in them that give the same equations of motion as the original local ones.  Such a procedure, even if confined to small finite variations in the location of a point from its partnered point, would be very interesting and might indicate a way to derive traditional nonlocal laws through their relation to local symmetries of the system.

\end{document}